\documentclass[aps,twocolumn]{revtex4}
\usepackage{amsmath}
\usepackage{graphicx}

\input{tcilatex}
\begin{document}

\title{A New Look At Gravitational Coupling Constant And The Dark Energy
Problem}
\author{O. F. Akinto$\footnotemark[1]$, Farida Tahir$\footnotemark[2]$}
\affiliation{Department of Physics, COMSATS Institute of Information Technology,
Islamabad, Pakistan}

\begin{abstract}
In this paper, we establish that the solution to the dark energy problem is
connected to the cutoff Ultraviolet $(UV)$ scale $M_{pl}$ manifesting itself
as linearly independent infrared sectors of the effective theory of gravity
interacting with $QCD$ fields. We work in the combined frameworks of finite
temperature - density corrections and effective quantum field theory (as low
energy quantum gravity). We strongly suggest that the failure to reproduce
the exact observed value of dark energy $\left( \rho _{\Lambda }\right) $
from the framework of Veneziano ghost theory of $QCD$ $\ $is intimately
linked to the unverifiable ad hoc assumption that conditions the
gravitational coupling constant to be unity $(C_{grav}=1).$ A close perusal
of the Minkowski vacuum structure reveals that $C_{grav}\neq 1.$We compute
the value of $C_{grav}$ from the Bose-Einstein distribution function. With $%
C_{grav}=1.797\times 10^{-1}$ coupled with the value of vacuum energy
estimated from the Veneziano ghost theory of $QCD$, we reproduce the
observed value of $\rho _{\Lambda }$ to be $\rho _{\Lambda }$ $\approx
C_{grav}$ $\left( 3.6\times 10^{-3}eV\right) ^{4}\approx \left( 2.3\times
10^{-3}eV\right) ^{4}$. An important prediction of these combined frameworks
(made manifest by the application of standard box-quantization procedure to
the $UV$ scale $M_{pl}$) states that there are $\approx 10^{122}$ linearly
independent \emph{\textquotedblleft subuniverses\textquotedblright\ }%
representing the linearly independent infrared sectors of the effective
theory of gravity interacting with $QCD$ fields. A direct consequence of
this is that our subuniverse is embedded on a non-trivial manifold M (such
as a torus group $T^{10^{122}}=T^{1}\times ..........\times T^{1}$) with
different linear sizes.

\end{abstract}

\maketitle

\section{Introduction}

\footnotetext[1]{%
pheligenius@yahoo.com}\footnotetext[2]{%
farida\_tahir@comsats.edu.pk}One of the most flummoxing problems in modern
physics that kept scientists at alert and has been hotly debated since $1929$%
, is the realization of the expansion of the universe, established when
Edwin Hubble published his revolutionary paper. Astronomical observations
and study of universe, in the past few decades, strongly invalidated
astronomers' view point that the universe was entirely composed of \emph{%
\textquotedblleft baryonic matter\textquotedblright }. The latest
conformation of the accelerating universe \cite{DN,AG,SP,SW,SE} endorsed the
fact that the universe is infused with an unknown form of energy density $($%
dubbed as dark energy $\left( \rho _{\Lambda }\right) )$ which makes up for
about $75\%$ of the total energy density of the universe. It is this $75\%$
mysterious $\rho _{\Lambda }$, which conditions our three-dimensional
spatial curvature to be zero, that is responsible for the acceleration of
the universe. This discovery provided the first direct evidence that $\rho
_{\Lambda }$ is non-zero, with $\rho _{\Lambda }\approx \left( 2.3\times
10^{-3}eV\right) ^{4}$\cite{FR,JZ}.

However, the theoretical expectations for the $\rho _{\Lambda }$ exceed
observational limits by some $120$ orders of magnitude \cite{JC}. This huge
discrepancy between theory and observation, hitherto, constitutes a serious
problem for theoretical physics community. In fact, Steven Weinberg puts it
more succinctly by saying that the small non-zero value of $\rho _{\Lambda }$
is \emph{\textquotedblleft a bone in the throat of theoretical
physics\textquotedblright .} \ Considering this huge discrepancy may
un-shroud something fundamental, yet to be unveiled, about the hidden nature
of the universe. This paper is one of such attempts.

The most elegant and comprehensible endeavour in order to solve this
problem, in our view, was put forward by F. R. Urban and A. R. Zhitnitsky 
\cite{FR}. These authors approached the problem from the angle of the
effective theory of gravity interacting with standard model fields by using
the solution of the $U(1)$ problem as put forward by G. Veneziano and E.
Witten \cite{GV,EW}.\ In this framework, the basic problem of why the dark
energy is $120$ orders of magnitude smaller than its Planck scale $%
M_{planck}^{4}$, is replaced by fundamentally different questions:
\textquotedblleft $(i)$ What is the relevant scale which enters the
effective theory of gravitation? $(ii)$ How does this scale appear in the
effective quantum field theory for gravity?\textquotedblright\ In their
view, this effective scale has nothing to do with the cutoff ultraviolet $%
(UV)$ scale $M_{planck}$: the appropriate effective scale must emerge as a
result of a subtraction at which some infrared (IR) scale enters the
physics. They completely turned the problem on its head!

Though their attempt being cognizant, yet it fails to reproduce, exactly,
the measured value of $\rho _{\Lambda }$ \cite{MT}. We observe here that
their assumption $g\equiv c=C_{QCD}\times C_{grav}=1$ is debatable, since it
is valid only for $C_{QCD}$ but not for $C_{grav}$, as proved in this paper.
Here $g$ is the Minkowski metric in vacuum, $C_{QCD}$ is the Quantum
Chromodynamic $(QCD)$ coupling constant, and $C_{grav}$ is the gravitational
coupling constant.

From Ref.[7], the value of $C_{grav}$ was wrongly computed to be $%
C_{grav}=0.0588$\ (which is approximately one-third of the value we proved
in our calculation) but for obvious reason the authors neglected this value
and used a position dependent Minkowski metric distance $g(x^{2})$\ instead.
They computed $g(x^{2})$ to be $g(x^{2})=1/6.25.$ For no clear reason, they
approximated the value of $g(x^{2})$ to $g(x^{2})\approx 1/6$ by truncating $%
0.25$ from their original \ value of $g(x^{2}).$ This approach is totally
unacceptable in the field of computational cosmology where every minuscule
value counts.

In this paper, we have proved the value of $C_{grav}$ to an order of
magnitude less than one i.e. $1.797\times 10^{-1}$; this leads towards the
exact measured \cite{MT} value of $\rho _{\Lambda }$. In order to get this
value, we have used finite temperature and density $(FTD)$ correction
technique. Here, the $FTD$ background acts as highly energetic medium $%
\left( M_{planck}^{4}\right) $ controlling the particle propagation. Our
basic guiding idea is that the finite temperature field theory $(FTFT)$,
similar to the physics of superconductivity (quantum field theory at $T=0$),
is linked to the infrared sector of the effective theory of gravity
interacting with standard model fields, specifically with $QCD$ fields \cite%
{FR}. In this case, the statistical background effects are incorporated in
propagators through the Bose-Einstein distribution function \cite{KR,FE}: it
is worth noting that the Bose-Einstein distribution function is the
mathematical tool for understanding the essential feature of the theory of
superconductivity \cite{FE}. The general attribute of a successful theory of
superconductivity is the existence of degenerate vacuum/broken symmetry
mechanism. A characteristic feature of such a theory is the possible
existence of \textquotedblleft unphysical\textquotedblright\ zero-mass
bosons which tend to preserve the underlying symmetry of the theory.\ The
masslessness of these singularities is protected in the limit $%
q\longrightarrow 0$. This means that it should cost no energy to create a
Yang-Mills quantum at $q=0$ and thus the mass is zero \cite{GS}. In the
preceding Ref. the Goldstone-Salam-Weinberg theorem is valid for a zero-mass
pole, which is protected. That pole is not physical and is purely gauge,
hence unphysical. This is precisely the highly celebrated Veneziano ghost 
\cite{FR}, which is analogous to the Kogut-Susskind \emph{(KS)} ghost in the
Schwinger model \emph{(distinctive unphysical degree of freedom which is
massless and can propagate to arbitrary large distances)}.

\qquad It is imperative to note that this set of unphysical massless bosons
tends to transform as a basis for a representation of a compact Lie group 
\cite{FE} thereby, forming a compact manifold. We do not make any specific
assumptions on the topological nature of the manifold; we only assume that
there is at least one Minkowski metric distance that defines a general
covariance of comoving coordinates \cite{SW} with size $L_{M}=2\times
Euclidean$ metric distance.

In the next section, we derive the finite temperature and density relation
for the Veneziano ghosts by using Bose-Einstein distribution function. It
should be noted here that Veneziano ghosts are treated as unphysical
massless bosons due to the fact that they both have the same propagator $%
(+ig_{\mu \nu }/q^{2})$\cite{FR}: the propagator for unphysical massless
boson is obtained from $(+\left( 2\pi \right) ^{4}ie^{2}g_{\mu \nu
}/q^{2})\langle \phi \rangle ^{2}$ \cite{FE}.

\section{ Veneziano-Ghost Density}

From Fermi-Dirac and Bose-Einstein distribution functions, we have%
\begin{equation}
n_{r}=\frac{g_{r}}{e^{\alpha +\beta \varepsilon _{r}}\pm 1}  \tag{1}
\end{equation}

The positive sign applies to fermions and the negative to bosons. $g_{r}$ is
the degenerate parameter, $\alpha $ is the coefficient of expansion of the
boson gas inside the volume $(V)$, $\beta $ is the Lagrange undetermined
multiplier, $n_{r}$ and $\varepsilon _{r}$ are the numbers of particles and
the energy of the $r-th$ state respectively. The value of $\alpha $ for
boson gas at a given temperature is determined by the normalization
condition \cite{PC}%
\begin{equation}
N=\underset{r}{\sum }\frac{g_{r}}{e^{\alpha +\beta \varepsilon _{r}}-1} 
\tag{2}
\end{equation}

This sum can be converted into an integral, because for a particle in a box,
the states of the system have been found to be very close together i.e. $%
\left( \Delta \varepsilon _{vac}\equiv d\varepsilon \rightarrow 0\right) $.
Using the density of single-particle states function, $Eq.(2)$ reduces to

\begin{equation}
N=\overset{\infty }{\underset{0}{\int }}\frac{D\left( \varepsilon \right)
d\varepsilon }{e^{\alpha +\beta \varepsilon }-1}  \tag{3}
\end{equation}

Where $D\left( \varepsilon \right) d\varepsilon $ is the number of allowed
states in the energy range $\varepsilon $ to $\varepsilon +d\varepsilon $
and $\varepsilon $ is the energy of the single-particle states. Using the
density of states as a function of energy, we have \cite{PC}

\begin{equation*}
D\left( \varepsilon \right) d\varepsilon =\frac{4\pi V}{h^{3}}2m\varepsilon
\left( \frac{m}{p}\right) d\varepsilon
\end{equation*}

with $p=\sqrt{2m\varepsilon }$

\begin{equation}
D\left( \varepsilon \right) d\varepsilon =2\pi V\left( \frac{2m}{h^{2}}%
\right) ^{3/2}\varepsilon ^{1/2}d\varepsilon  \tag{4}
\end{equation}

Putting $Eq.(4)$ into $Eq.(3),$ we get

\begin{equation}
N=2\pi V\left( \frac{2m}{h^{2}}\right) ^{3/2}\overset{\infty }{\underset{0}{%
\int }}\frac{\varepsilon ^{1/2}d\varepsilon }{e^{\alpha +\beta \varepsilon
}-1}  \tag{5}
\end{equation}

Where $m$ is the mass of boson and $h$ is the Planck constant. $\alpha
=\beta \mu $ and $\beta =1/kT$. $\mu $ is the chemical potential, $k$ is the
Boltzmann constant and $T$ is temperature. Since there is no restriction on
the total number of bosons, the chemical potential is always equals to zero.
Thus $Eq.(5)$ reads as:

\begin{equation}
N=2\pi V\left( \frac{2m}{h^{2}}\right) ^{3/2}\overset{\infty }{\underset{0}{%
\int }}\frac{\varepsilon ^{1/2}d\varepsilon }{e^{\varepsilon /kT}-1}  \tag{6}
\end{equation}

By using standard integral

\begin{equation*}
\overset{\infty }{\underset{0}{\int }}\frac{x^{z-1}dx}{e^{x}-1}=\varsigma
\left( z\right) \Gamma \left( z\right)
\end{equation*}

where $\varsigma \left( z\right) $ is the Riemann zeta function and $\Gamma
\left( z\right) $ is the gamma function. $Eq.(6)$ takes the form

\begin{equation*}
N=2.61V\left( \frac{2\pi mkT}{h^{2}}\right) ^{3/2}
\end{equation*}

Let $n_{gv}=N/V$

\begin{equation}
n_{gv}=2.61\left( \frac{2\pi mkT}{h^{2}}\right) ^{3/2}  \tag{7}
\end{equation}

\bigskip Recall that \ $m=\Delta \varepsilon _{vac}/c^{2}$ and the average
kinetic energy of gas in three-dimensional space is given by $\Delta
\varepsilon _{vac}=\frac{3kT}{2}$. Thus $Eq.(7)$ becomes

\begin{equation*}
n_{gv}=\left( \frac{\left( 2.61\right) \left( 3\pi \right) ^{3/2}k^{3}}{%
\left( hc\right) ^{3}}\right) T^{3}
\end{equation*}

Define

\begin{eqnarray*}
\xi &\equiv &\left( \frac{\left( 2.61\right) \left( 3\pi \right) ^{3/2}k^{3}%
}{\left( hc\right) ^{3}}\right) \\
&=&2.522\times 10^{7}\left( mk\right) ^{-3}
\end{eqnarray*}

Hence, the Veneziano-ghost density$\left( n_{gv}\right) $ can be
re-expressed in more elegant form as:

\begin{equation}
n_{gv}=\xi T^{3}  \tag{8}
\end{equation}

\bigskip $Eq.(8)$ is the required result for the finite temperature and
density relation for the Veneziano ghost(s).

\section{Gravitational Coupling Constant From Veneziano-Ghost Density}

The principle of general covariance tells us that the energy-momentum tensor
in the vacuum must take the form

\begin{equation}
\left\langle 0\left\vert \widehat{T}_{\mu \nu }\right\vert 0\right\rangle
=T_{\mu \nu }^{vac}=g\left\langle \rho \right\rangle  \tag{9}
\end{equation}

\bigskip Here $\left\langle \rho \right\rangle $ has the dimension of energy
density and $g$ describes a real gravitational field \cite{SE}. Thus $Eq.(9)$
can be written as

\begin{equation}
\left\langle 0\left\vert \widehat{T}_{\mu \nu }\right\vert 0\right\rangle
=g\left( \Delta \varepsilon _{vac}\right) ^{4}  \tag{10}
\end{equation}

Where \textquotedblleft $g$\textquotedblright\ in Ref.\cite{FR,JZ}, is
defined as $g\equiv c=C_{QCD}\times C_{grav}.$ Therefore, $Eq.(10)$ can be
written as

\begin{equation*}
\left\langle 0\left\vert \widehat{T}_{\mu \nu }\right\vert 0\right\rangle
=C_{QCD}\times C_{grav}\times \left( \Delta \varepsilon _{vac}\right) ^{4}
\end{equation*}

Where, $C_{QCD}=1$ as quoted by \cite{JZ}, and references within, thus

\begin{equation}
\left\langle 0\left\vert \widehat{T}_{\mu \nu }\right\vert 0\right\rangle
=C_{grav}\times \left( \Delta \varepsilon _{vac}\right) ^{4}  \tag{11}
\end{equation}

Now, the energy density can be written as

\begin{equation}
\rho _{vac}=\frac{\Delta \varepsilon _{vac}}{V}=V^{-1}\times \Delta
\varepsilon _{vac}  \tag{12}
\end{equation}

$Eq.(12)$ is justified by the standard box-quantization procedure \cite{SE}.
By comparing $Eq.(12)$ with $Eq.(8)$, we get

\begin{equation}
\rho _{vac}=n_{gv}\times \Delta \varepsilon _{vac}  \tag{13}
\end{equation}

With $n_{gv}\equiv V^{-1},$ From the average kinetic energy for gas in
three-dimensional space, we have $T=2\Delta \varepsilon _{vac}/3k.$ Hence $%
Eq.(8)$ becomes

\begin{equation}
n_{gv}=\frac{8\xi \left( \Delta \varepsilon _{vac}\right) ^{3}}{27k^{3}} 
\tag{14}
\end{equation}

Putting the value of $n_{gv}$ in $Eq.(13)$, we get

\begin{equation}
\rho _{vac}=\frac{8\xi \left( \Delta \varepsilon _{vac}\right) ^{4}}{27k^{3}}
\tag{15}
\end{equation}

$Eq.(15)$ represents the energy density of a vacuum state.

The natural demand of the Lorentz invariance of the vacuum state is bedecked
in the structure of (effective) quantum field theory in Minkowski space-time
geometry \cite{SE,RM}. Hence, if $\left\vert 0\right\rangle $\ is a vacuum
state in a reference frame $S$ and $\left\vert 0%
{\acute{}}%
\right\rangle $ refers to the same vacuum state observed from a reference
frame $S%
{\acute{}}%
,$\ which moves with uniform velocity relative to $S$, then the quantum
expression for Lorentz invariance of the vacuum state reads%
\begin{equation}
\left\vert 0%
{\acute{}}%
\right\rangle =u\left( L\right) \left\vert 0\right\rangle =\left\vert
0\right\rangle  \tag{16}
\end{equation}

Where $u\left( L\right) $ is the unitary transformation (acting on the
quantum state $\left\vert 0\right\rangle $) corresponding to a Lorentz
transformation $L$. All the physical properties that can be extracted from
this vacuum state, such as the value of energy density, should also remain
invariant under Lorentz transformations \cite{SE}. If the Lorentz
transformation is initiated by $\rho _{vac}$, then $2\times \rho _{vac}$ is
needed for a unitary transformation to take place. The logic behind this
assumption is simple: if $\rho _{vac}$ defines the Lorentz invariant length $%
(L)$ (Euclidean metric distance) of $\left\vert 0\right\rangle $, then the
Lorentz transformation from $\left\vert 0\right\rangle $ to $\left\vert 0%
{\acute{}}%
\right\rangle $ (with continuous excitation) requires \ $2\times \rho _{vac}$%
: \ $\left\vert 0\right\rangle \overset{2\times \rho _{vac}}{\longrightarrow 
}$\ $\left\vert 0%
{\acute{}}%
\right\rangle $. This leads to the principle of general covariance as
apriori stated in the introduction \cite{SE}. Thus,

\begin{equation}
\left\langle 0\left\vert \widehat{T}_{\mu \nu }\right\vert 0\right\rangle
=2\times \rho _{vac}=\frac{16\xi \left( \Delta \varepsilon _{vac}\right) ^{4}%
}{27k^{3}}  \tag{17}
\end{equation}

\bigskip $Eq.(17)$ is also justified by the standard box-quantization
procedure \cite{SE}. Now by combining $Eq.(11)$ and $Eq.(17)$, we have

\begin{equation*}
C_{grav}=\frac{16\xi }{27k^{3}}=2.336\times 10^{19}\left( m.eV\right) ^{-3}
\end{equation*}

\bigskip As $1m=5.07\times 10^{15}GeV^{-1}$. This leads to

\begin{equation}
C_{grav}=1.797\times 10^{-1}  \tag{18}
\end{equation}

which is the required gravitational coupling constant.

\section{Dark Energy From The Veneziano-Ghost: A Review}

The major ingredient of standard Witten-Veneziano resolution of $U(1)$
problem is the existence of topological susceptibility $\chi $. In Ref.\cite%
{FR}, it has been proved that the deviation in $\chi $, i.e. $\Delta \chi $,
represents the vacuum energy density \emph{(dark energy)}. We review this
result by making use of $Eq.(9)$ and resolve the inherent hitch in this
approach with the help of $Eq.(18)$. Thus from $Eq.(9)$ we have,

\begin{equation}
i\int dx\left\langle 0\left\vert \widehat{T}_{\mu \nu }\right\vert
0\right\rangle =i\int dxT_{\mu \nu }^{vac}  \tag{19}
\end{equation}

By using the standard Witten-Veneziano relations

\begin{equation*}
\widehat{T}_{\mu \nu }\equiv T\left\{ Q\left( x\right) ,Q\left( 0\right)
\right\}
\end{equation*}

\bigskip Where%
\begin{equation*}
Q\equiv \frac{\alpha _{s}}{16\pi }\epsilon ^{\mu \nu \rho \sigma }G_{\mu \nu
}^{a}G_{\rho \sigma }^{a}\equiv \frac{\alpha _{s}}{8\pi }G_{\mu \nu }^{a}%
\widetilde{G}^{\mu \nu a}\equiv \partial _{\mu }K^{\mu }
\end{equation*}

And

\begin{equation}
K^{\mu }\equiv \frac{\Gamma ^{2}}{16\pi ^{2}}\epsilon ^{\mu \nu \lambda
\sigma }A_{\nu }^{a}\left( \partial _{\lambda }A_{\sigma }^{a}+\frac{\Gamma 
}{3}f^{abc}A_{\lambda }^{b}A_{\sigma }^{c}\right)  \tag{20}
\end{equation}

\bigskip Where $A_{\mu }^{a}$\ are the conventional $QCD$ color gluon fields
and $Q$ is the topological charge density, and $\alpha _{s}=$ $\frac{\Gamma
^{2}}{4\pi }$. Thus we have

\begin{equation*}
i\int dx\left\langle 0\left\vert T\left\{ Q\left( x\right) ,Q\left( 0\right)
\right\} \right\vert 0\right\rangle =i\int dxT_{\mu \nu }^{vac}
\end{equation*}

\begin{equation}
\underset{q\longrightarrow 0}{\lim }i\int dxe^{iqx}\left\langle 0\left\vert
T\left\{ Q\left( x\right) ,Q\left( 0\right) \right\} \right\vert
0\right\rangle =\underset{q\longrightarrow 0}{\lim }i\int dxe^{iqx}T_{\mu
\nu }^{vac}  \tag{21}
\end{equation}

\bigskip Let%
\begin{equation*}
\underset{q\longrightarrow 0}{\lim }i\int dxe^{iqx}T_{\mu \nu }^{vac}=\chi
\end{equation*}

Hence $Eq.(21)$ becomes

\begin{equation*}
\chi =\underset{q\longrightarrow 0}{\lim }i\int dxe^{iqx}\left\langle
0\left\vert T\left\{ Q\left( x\right) ,Q\left( 0\right) \right\} \right\vert
0\right\rangle
\end{equation*}

And

\begin{equation}
\Delta \chi =\Delta \left[ \underset{q\longrightarrow 0}{\lim }i\int
dxe^{iqx}\left\langle 0\left\vert T\left\{ Q\left( x\right) ,Q\left(
0\right) \right\} \right\vert 0\right\rangle \right]  \tag{22}
\end{equation}

Using $\Delta =c\left( H/m_{\eta }\right) $ and $\left[ \underset{%
q\longrightarrow 0}{\lim }i\int dxe^{iqx}\left\langle 0\left\vert T\left\{
Q\left( x\right) ,Q\left( 0\right) \right\} \right\vert 0\right\rangle %
\right] =-\left[ \lambda _{YM}^{2}\left( q^{2}-m_{0}^{2}\right) /\left(
q^{2}-m_{0}^{2}-\frac{\lambda _{\eta }^{2}}{N_{c}}\right) \right] $ from \
Ref.\cite{FR}, $Eq.(22)$ can be written as

\begin{equation}
\Delta \chi =-c\left( \frac{2H}{m_{\eta }}\right) .\frac{\lambda
_{YM}^{2}\left( q^{2}-m_{0}^{2}\right) }{\left( q^{2}-m_{0}^{2}-\frac{%
\lambda _{\eta }^{2}}{N_{c}}\right) }  \tag{23}
\end{equation}

The standard Witten-Veneziano solution of $U(1)$ problem is based on the
well-established assumption (confirmed by various lattice computations) that
\ $\chi $ does not vanish, despite the fact that $Q$ is a total derivative $%
Q\equiv \partial _{\mu }K^{\mu }$. This suggests that there is an unphysical
pole at $q=0$ in the correlation function of $K^{\mu }$, similar to \emph{KS}
ghost in the Schwinger model \cite{FR}. Thus $Eq.(23)$ becomes

\begin{equation}
\Delta \chi =-c\left( \frac{2H}{m_{\eta }}\right) .\frac{\lambda
_{YM}^{2}m_{0}^{2}}{m_{\eta }^{2}}  \tag{24}
\end{equation}

where $m_{\eta }^{2}=m_{0}^{2}+\frac{\lambda _{\eta }^{2}}{N_{c}}$ is the
mass of physical $\eta $ field and the reason for a factor of $2$ in $%
Eq.(24) $ follows from the principle of general covariance as we have
already established. Using Witten-Veneziano relation \ $4\lambda _{YM}^{2}$\ 
$=f_{\pi }^{2}m_{\eta }^{2}$ and chiral condensate $m_{0}^{2}f_{\pi
}^{2}=-4m_{q}\left\langle \overline{q}q\right\rangle ,$ \ $Eq.(24)$ \ can be
written as

\begin{equation}
\Delta \chi =c\left( \frac{2H}{m_{\eta }}\right) \left\vert
m_{q}\left\langle \overline{q}q\right\rangle \right\vert  \tag{25}
\end{equation}

where $H$ is Hubble constant and $m_{q}$\ is the mass of a single light
quark. From Ref.\cite{FR} $c\left( \frac{2H}{m_{\eta }}\right) \left\vert
m_{q}\left\langle \overline{q}q\right\rangle \right\vert \approx c\left(
3.6\times 10^{-3}eV\right) ^{4}$ leads to

\begin{equation}
\Delta \chi \approx c\left( 3.6\times 10^{-3}eV\right) ^{4}  \tag{26}
\end{equation}

By using $c=C_{QCD}\times C_{grav}\approx C_{grav}$ from \cite{JZ} and
reference within, $Eq.(26)$ can be written as

\begin{equation}
\Delta \chi \approx C_{grav}\left( 3.6\times 10^{-3}eV\right) ^{4}  \tag{27}
\end{equation}

\bigskip Comparision of $Eq.(18)$ with $Eq.(27)$ gives 
\begin{equation}
\rho _{\Lambda }\equiv \Delta \chi \approx \left( 2.3\times 10^{-3}eV\right)
^{4}  \tag{28}
\end{equation}

$Eq.(28)$ is the measured value of $\rho _{\Lambda }$ that is responsible
for the acceleration of the universe.

\ \ \ \ \ \ \ \ \ \ \ \ \ Using Planck scale $M_{Pl}^{4}$ as the cutoff
correction, $Eq.(8)$ becomes

\begin{equation}
n_{gv}^{planck}=7\times 10^{103}m^{-3}  \tag{29}
\end{equation}

From the standard box-quantization procedure \cite{SE}, we have

\begin{equation}
2\times \rho _{\Lambda }^{total}=\frac{1}{V}\underset{k}{\sum \text{%
h{\hskip-.2em}\llap{\protect\rule[1.1ex]{.325em}{.1ex}}{\hskip.2em}%
}\omega _{k}}  \tag{30}
\end{equation}

By imposing Lorentz invariance of vacuum state formalism on $Eq.(30)$, we
have

\begin{equation}
2\times \rho _{\Lambda }^{total}=\frac{1}{V}n\text{%
h{\hskip-.2em}\llap{\protect\rule[1.1ex]{.325em}{.1ex}}{\hskip.2em}%
}\omega =n\left[ n_{gv}\times \Delta \varepsilon _{vac}\right]  \tag{31}
\end{equation}

Where $n_{gv}\equiv \frac{1}{V}$ and $\Delta \varepsilon _{vac}\equiv $%
h{\hskip-.2em}\llap{\protect\rule[1.1ex]{.325em}{.1ex}}{\hskip.2em}%
$\omega .$ Note that $Eq.(31)$ reduces to $Eq.(17)$ for $n=1,$ therefore $%
Eq.(31)$ can be rewritten for Planck scale cutoff correction (where Planck
series of energy $(n$%
h{\hskip-.2em}\llap{\protect\rule[1.1ex]{.325em}{.1ex}}{\hskip.2em}%
$\omega )$ is taken to be the Planck energy $(E_{Pl})$):

\begin{equation}
n\left[ n_{gv}\times \Delta \varepsilon _{vac}\right] =n_{gv}^{Planck}\times
E_{Pl}  \tag{32}
\end{equation}

From $Eqs.(13),(14),(15),(17)$ and $(28),$ we have

\begin{equation*}
n\left[ \frac{8\xi \left( \Delta \varepsilon _{vac}\right) ^{4}}{27k^{3}}%
\right] =n_{gv}^{Planck}\times E_{Pl}
\end{equation*}

\begin{equation}
n\left[ \frac{\rho _{\Lambda }}{2}\right] =n_{gv}^{Planck}\times
E_{Pl}=M_{Pl}^{4}  \tag{33}
\end{equation}

Where $\rho _{vac}=\rho _{\Lambda }/2$ is the energy density of each
infrared sector. $Eq.(33)$ shows how cutoff $UV$ scale $M_{Pl}^{4}$\
manifests itself as linearly independent infrared sectors of the effective
theory of gravity interacting with $QCD$ fields.

By combining $Eqs.(28),(29)$ and $(33)$ we have

\begin{equation}
n=4\times 10^{122}\approx 10^{122}  \tag{34}
\end{equation}

Where $n_{gv}^{Planck}\times E_{Pl}=M_{Pl}^{4}=1.4\times 10^{113}J/m^{3}.$
Thus $Eq.(34)$ suggests that there are $\approx 10^{122}$(degenerate) vacuum
states. These vacuum states ($n-$torus) are called \emph{\textquotedblleft
subuniverses or multiverse"\ }\cite{EB,SWH,SC,AV,SW97}. An $n-$torus is an
example of $n-$dimensional compact manifold or a compact Abelian Lie group $%
U(1).$ In this sense, it is a product of $n$ circles i.e $T^{n}=S^{1}\times
S^{1}\times ..........\times S^{1}=T^{1}\times T^{1}\times ........\times
T^{1}$ \cite{BR,DL,RP}. In this paper, $n$ circles, which are the elements
of $U(1)$ group, represent $n$ linearly independent infrared sectors or the
unphysical massless gauge bosons dubbed as Veneziano ghosts.

It is important to notice that the existence of non-vanishing \ and linearly
independent infrared sectors of the effective theory of gravity interacting
with $QCD$ fields is parametrically proportional to the Planck cutoff
energy. Therefore, our simple extension of Veneziano ghost theory of $QCD$
to accommodate FTFT has striking consequences: it predicts, accurately, the
value of \ $C_{grav},$ which leads towards the $100\%$ consistency between
theory and experimental value of $\rho _{\Lambda }.$ As an offshoot, it
fotifies the idea of multiverse and paints a new picture of quantum
cosmological paradigm.

\section{Summary and Conclusion}

The computational analysis of the dark energy problem from the combined
frameworks of finite temperature-density correction technique and the
Veneziano ghost theory of $QCD$ conditions $FTD$ background to behave like a
reservoir for the infrared sectors of the effective theory of gravity
interacting with $QCD$ fields. These infrared sectors (unphysical massless
bosons) transform as a basis for a representation of a compact manifold.
This is analogous to the process of quantizing on manifold $M$ (such as a
torus group $T^{n}=T^{1}\times ..........\times T^{1}$ $=T^{10^{122}}$), in
which all the submanifolds (tori) are linearly independent of each other.
This means that an \emph{\textquotedblleft observer\textquotedblright }\
trapped in one of such tori would think his torus is the whole Universe. An
important prediction of this is that the vacuum energy $\Delta \varepsilon
_{vac}$ owes its existence to the degenerate nature of vacuum (or to the
asymmetric nature of the universe). The effect of this is a direct
consequence of the embedding of our subuniverse on a non-trivial manifold $M$
with (minuscule) different linear sizes.

The main result of the present study is that the effective scales obviously
have something to do with the cutoff Ultraviolet $(UV)$ scale $M_{Pl}$.
Based on the standard box-quantization procedure, the $UV$ scale $M_{Pl}$ is
a collection of infrared $(IR)$ scales. Undoubtedly, the relevant effective
scales appear as a result of energy differences (subtractions) at which the $%
IR$ scales enter the physics of $UV$ scale $M_{Pl}$. It is therefore
impossible to compute the value of $\rho _{\Lambda }$ without putting into
consideration the statistical effect of the $UV$ scale $M_{Pl}$ which
manifests itself through the existence of the linearly independent $IR$
sectors of the effective theory of quantum field theory $(QFT)$: this is the 
\emph{\textquotedblleft stone\textquotedblright } that confirms the
interrelationship between $FTFT$ and the theory of superconductivity $(QFT$
at $T=0)$.

Thus, if you buy the idea of Lorentz invariance of vacuum state formalism or
the degenerate vacuum mechanism, then $\sim 10^{122}$ subuniverses come as
free gifts!

\subsubsection{\textbf{Acknowledgement}}

Mr. O. F. Akinto is indebted to the Department of Physics, CIIT, Islamabad \
and the National Mathematical Center Abuja, Nigeria \ for their finacial
support.

\newpage


\bigskip

\begin{thebibliography}{99}
\bibitem{DN} D. N. Spergel, et al., WMAP Collaboration, Astrophys. J. Suppl.
148 (2003) 175.

\bibitem{AG} A. G. Riess, et al., Supernova Search Team Collaboration,
Astron. J. 116 (1998).

\bibitem{SP} S. Perlmutter, et al., Supernova Cosmology Project
Collaboration, Astrophys. J. 515 (1999) 565.

\bibitem{SW} S. Weinberg, Reviews of Mod. Phys. Vol. 61, No 1 (1998).

\bibitem{SE} S. E. Rugh \& H. Zinkernagel, Studies in History and Philosophy
of Science part B 33 (2002) 4.

\bibitem{FR} F. R. Urban \& A. R. Zhitnitsky, Phy. Lett. B 688 (2010) 9.

\bibitem{JZ} J. Zhou et al., Mod. Phys. 1172 (2012) 3.

\bibitem{JC} E. J. Copeland et al., Int. J. Mod. Phys. D15 (2006).

\bibitem{GV} G. Veneziano, Nucl. Phys. B 159 (1979) 213.

\bibitem{EW} E. Witten, Nucl. Phys. B 156 (1979) 269.

\bibitem{MT} M. Tegmark, M. A. Strauss, M. R. Blanton, et al., Phys. Rev. D
69 (2004)103501.

\bibitem{KR} K. Raja opal \& F. Wilczek, The condensed Matter physics of
QCD, hep-ph/0103141.

\bibitem{FE} F. Englert \& R. Brout, Phys. Rev. Lett.13 (1964) 9.

\bibitem{GS} G. S. Guralnik, Proceedings of the DPF- Guy Wilkinson,
hep-ex/0811.2642. Conference, Providence, RI (2011)

\bibitem{PC} P. C. Riedi, \textquotedblleft Thermal
Physics\textquotedblright , Macmillan Press Ltd, 1st ed.(1976).

\bibitem{RM} R. M. Wald, \textquotedblleft Quantum Field Theory in Curved
Space-time and Black Hole Thermodynamics, The University\ Of Chicago Press
(1994).

\bibitem{EB} E. Baum, Phys. Lett. B133 (1984) 185.

\bibitem{SWH} S. W. Hawking, Phys. Lett. B134 (1984) 403.

\bibitem{SC} S. Coleman, Nucl. Phys. B307 (1988) 867.

\bibitem{AV} A. Vilenkin, Phys. Rev. D27 (1983) 2848.

\bibitem{SW97} S. Weinberg et al., astro-ph/9701099.

\bibitem{BR} B. Reinhold, Duke Mathematical Journal 3(1937) 1.

\bibitem{DL} D.L. Johnson, " Symmetries", Springer, (2001). \ 

\bibitem{RP} R.Penrose, "The Road to Reality", Vintage Books Ltd, 3rd
Edition, (2004).
\end{thebibliography}
\end{document}